\documentclass[superscriptaddress,twocolumn,longbibliography,amsmath,amssymb,floatfix,aps]{revtex4-1}
\usepackage[dvipdfmx]{graphicx,color}
\usepackage{dcolumn}
\usepackage{bm}
\usepackage{here}

\begin{document}

\preprint{AIP/123-QED}

\title[Sample title]{Role of typical elements in Nd$_{2}$Fe$_{14}X$ ($X$ = B, C, N, O, F)}%

\author{Yasutomi Tatetsu}
\email{tatetsu.y.aa@m.titech.ac.jp}
\affiliation{Department of Materials Science and Engineering, Tokyo Institute of Technology, Yokohama 226-8502, Japan}
\affiliation{ESICMM, National Institute for Materials Science, Tsukuba 305-0047, Japan.}%

\author{Yosuke Harashima}
\email{yosuke-harashima@aist.go.jp}
\affiliation{ESICMM, National Institute for Materials Science, Tsukuba 305-0047, Japan.}%
\affiliation{CD-FMat, National Institute of Advanced Industrial Science and Technology, Tsukuba 305-8568, Japan}

\author{Takashi Miyake}%
\affiliation{ESICMM, National Institute for Materials Science, Tsukuba 305-0047, Japan.}%
\affiliation{CD-FMat, National Institute of Advanced Industrial Science and Technology, Tsukuba 305-8568, Japan}
\affiliation{CMI$^2$, MaDIS, National Institute for Materials Science, Tsukuba 305-0047, Japan}

\author{Yoshihiro Gohda}
\affiliation{Department of Materials Science and Engineering, Tokyo Institute of Technology, Yokohama 226-8502, Japan}%
\affiliation{ESICMM, National Institute for Materials Science, Tsukuba 305-0047, Japan.}%

\date{\today}

\begin{abstract}
The magnetic properties and structural stability of Nd$_{2}$Fe$_{14}X$ ($X$ = B, C, N, O, F) 
are theoretically studied by first-principles calculations focusing on the role of $X$.
We find that B reduces the magnetic moment (per formula unit) and magnetization (per volume) in Nd$_{2}$Fe$_{14}$B. 
The crystal-field parameter $A_2^0 \langle r^2 \rangle$ of Nd is not enhanced either, 
suggesting that B has minor roles in the uniaxial magnetocrystalline anisotropy of Nd.
These findings are in contrast to the long-held belief that 
B works positively for the magnetic properties of Nd$_{2}$Fe$_{14}$B. 
As $X$ changes from B to C, N, O and F, both the magnetic properties and stability vary significantly. 
The formation energies of Nd$_{2}$Fe$_{14}X$ and $\alpha$-Fe 
relative to that of Nd$_{2}$Fe$_{17}X$ are negative for $X$ = B and C, whereas they are positive when $X$ = N, O and F. 
This indicates that B plays an important role in stabilizing the Nd$_{2}$Fe$_{14}$B phase. 
\end{abstract}

\pacs{Valid PACS appear here}
\maketitle

\section{Introduction}
\label{introduction}
Back in the 1970's, Sm$_2$Co$_{17}$ was the strongest known hard magnetic compound. 
As Co is an expensive element whereas Fe is relatively cheap, 
developing iron-based permanent magnets was required.
Sm$_{2}$Fe$_{17}$, more generally known as R$_{2}$Fe$_{17}$ where R represents rare-earth elements, was a natural choice.
However, the Curie temperature $T_{c}$ of R$_{2}$Fe$_{17}$ is too low for practical applications.
For example, $T_{c}$ of Nd$_{2}$Fe$_{17}$ is about 330 K~\cite{Nd2Fe17-3, Nd2Fe17-1, Nd2Fe17-2}.
Sagawa came up with the following idea in order to raise the $T_{c}$ of R$_{2}$Fe$_{17}$:
The introduction of light elements, such as B and C, at the interstitial region in R$_{2}$Fe$_{17}$, would expand the volume of R$_{2}$Fe$_{17}$~\cite{Sagawa1, Sagawa-http}.
This volume expansion could make the ferromagnetism stronger by the magnetovolume effect, consequently $T_{c}$ would go up. 
He added B to Nd$_{2}$Fe$_{17}$ based on this idea, and successfully developed Nd-Fe-B magnets having a $T_{c}$ of 585 K~\cite{Sagawa1, Sagawa2}.
It is important to note that the main phase of Nd-Fe-B magnets is Nd$_{2}$Fe$_{14}$B, not Nd$_{2}$Fe$_{17}$B.

After the Nd-Fe-B magnet was developed, the role of B on the magnetism of Nd$_{2}$Fe$_{14}$B was studied theoretically~\cite{Kanamori1, Kanamori2, Kanamori3}. 
Kanamori pointed out the importance of the $p$-$d$ hybridization between B and Fe sites. 
The magnetic moment of Fe in the vicinity of B should be suppressed. It is called cobaltization. 
The cobaltized Fe (pseudo Co) can enhance the magnetic moment of the surrounding Fe 
due to the $d$-$d$ hybridization between pseudo Co and Fe, as is in Fe-Co alloys ~\cite{Hasegawa,Hamada}.
It has been believed for decades that when these chemical effects, owing to hybridizations, are summed up, the total magnetic moment of Nd$_{2}$Fe$_{14}$B increases by B.

Asano and Yamaguchi studied the magnetic properties of Y$_{2}$Fe$_{14}$B and the 
hypothetical compound Y$_{2}$Fe$_{14}$ through first-principles calculations ~\cite{Asano1, Asano2}. 
Their results show 
(i) the magnetic moment of Fe, at the nearest neighbor of B in Y$_{2}$Fe$_{14}$B, is reduced by the cobaltization, 
(ii) the magnetic moment of Fe at the nearest neighbor of pseudo Co is enhanced, and 
(iii) the effect of B is slightly negative on the magnetization in Y$_{2}$Fe$_{14}$B. 
Therefore, Y$_{2}$Fe$_{14}$B has smaller total magnetic moment than Y$_{2}$Fe$_{14}$. 
However, they optimized only lattice parameters within the atomic-sphere approximation, and internal coordinates were fixed. 
Moreover, magnetovolume effects and chemical effects were not distinguished. 
Therefore, the effects of B on the magnetism in R$_{2}$Fe$_{14}$B still needs to be clarified.

In this study, we systematically investigate the influence of B on the electronic states and magnetism of Nd$_{2}$Fe$_{14}$B through first-principles calculations. 
We find that the total magnetic moment (per formula unit) and magnetization (per volume) of Nd$_{2}$Fe$_{14}$B decrease by the addition of B . 
This result is analyzed in terms of the chemical effect and magnetovolume effect caused by B. 
The local magnetic moment at each Fe site is computed and is discussed in connection with cobaltization. 
We also study Nd$_{2}$Fe$_{14}X$, where $X$ represents C, N, O and F, in order to investigate the effects of the $X$ element on the magnetization, magnetocrystalline anisotropy and stability of Nd$_{2}$Fe$_{14}X$. 

\section{Computational methods}
We perform first-principles calculations of Nd$_{2}$Fe$_{14}X$ within density functional theory. 
We use the computational package OpenMX, which is based on pseudopotentials and pseudo-atomic-orbital basis functions~\cite{OpenMX}. 
The basis set of Nd, Fe and $X$ is $s2p2d2$, and the cutoff radii of these atoms are 8.0, 6.0 and 7.0 a.u., respectively.
For semicore states, we treat 3$s$ and 3$p$ orbitals as valence electrons in Fe, as well as 5$s$ and 5$p$ in Nd. 
An open-core pseudopotential is used for Nd atoms, where 4$f$ electrons are treated as spin-polarized core electrons.
The calculation for valence electronic states does not include the spin-orbit interaction. 
We adopt the Perdew-Burke-Ernzerhof exchange-correlation functional in the generalized gradient approximation~\cite{GGA}.
The lattice parameters and atomic positions of Nd$_{2}$Fe$_{14}X$ are all relaxed. 
For the convergence criteria, the maximum force on each atom and the total energy are 10$^{-4}$ hartree/bohr and 10$^{-7}$ hartree, respectively.
Spin collinear structures are assumed in all the calculations for Nd$_{2}$Fe$_{14}X$.
The cutoff energy is 500 Ry, and 8 $\times$ 8 $\times$ 6 $k$-point meshes are adopted for all the calculations. 

The magnetic moment is estimated from the spin moment of valence electrons and the contribution of Nd-4$f$ electrons $g_{J}J$ = 3.273 $\mu_{\text{B}}$, 
where $g_{J}$ is the Lande $g$-factor and $J$ is the total angular momentum of Nd-4$f$ electrons. 
The magnetization is calculated from the total magnetic moment of Nd$_{2}$Fe$_{14}X$ by dividing volume $V$ (see Appendix~\ref{app:latticeconstants}) and multiplying the Bohr magneton $\mu_{\text{B}}$.
The crystal-field parameters $A_{l}^{m}$ of Nd are calculated by using the computational package QMAS, which is based on 
the projector augmented-wave method~\cite{QMAS}.
See Refs.~\cite{Miyake,Harashima2} for more details.

We note that there are different notations for the Wyckoff positions of Nd$_{2}$Fe$_{14}$B~\cite{Herbst1,Shoemaker,Givord,Herbst2}.
Throughout this paper, the notation given in Refs.~\onlinecite{Herbst1,Herbst2} is used.

\section{Results and Discussion}
\subsection{Roles of B in Nd$_{2}$Fe$_{14}$B}
\begin{table}[b]
  \caption{The magnetic moment $m$ [$\mu_{\text{B}}$/f.u.] and magnetization $\mu_{0}M$ [T] 
  of Nd$_{2}$Fe$_{14}$B, Nd$_{2}$Fe$_{14}$B$_{0}$, and Nd$_{2}$Fe$_{14}$.}
  \vspace{5pt}
  \begin{tabular}{lcc}
    \hline \hline
    &  $m$ [$\mu_{\text{B}}$/f.u.] & $\mu_{0}M$ [T] 
    \\
    \hline
    Nd$_{2}$Fe$_{14}$B & 37.42 & 1.86
    \\
    Nd$_{2}$Fe$_{14}$B$_{0}$ & 38.87 & 1.93
    \\
    Nd$_{2}$Fe$_{14}$ & 38.43 & 1.95
    \\ [2pt]
    \hline \hline
  \end{tabular}
  \label{table:mag_nd2fe14b}
\end{table}

\begin{figure}[ht]
\includegraphics[width=8.5cm]{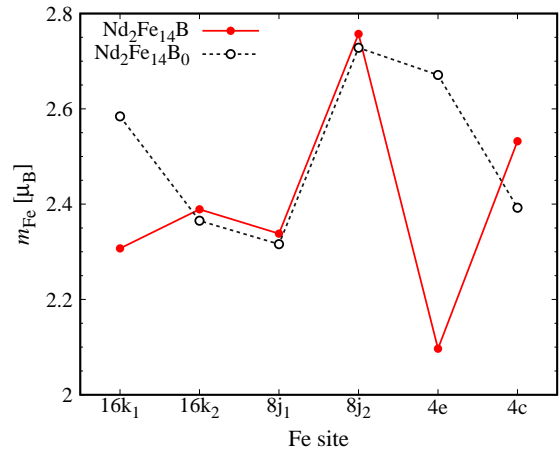}
\caption{(Color online) The local magnetic moments of Fe at each site $m_{\text{Fe}}$ in Nd$_{2}$Fe$_{14}$B (closed red circles) and Nd$_{2}$Fe$_{14}$B$_{0}$ (open black circles). 
  The site notation follows Refs.~\onlinecite{Herbst1,Herbst2}.
  Lines are provided as a visual guide.}
\label{fig:localmoment_nd2fe14b}
\end{figure}

We calculate the magnetic moment and magnetization of Nd$_{2}$Fe$_{14}$B and Nd$_{2}$Fe$_{14}$ in order to investigate the effects of B on Nd$_{2}$Fe$_{14}$B. 
Nd$_{2}$Fe$_{14}$ is a hypothetical compound in which the $X$ site is empty and all the atomic positions and lattice parameters are optimized.
For comparison, we also study Nd$_{2}$Fe$_{14}$B$_{0}$, 
in which the lattice parameters and atomic positions are the same as those of Nd$_{2}$Fe$_{14}$B, but B has been removed. 
These hypothetical compounds are introduced in order to distinguish the magnetovolume and chemical effects on the magnetic moment and magnetization of Nd$_{2}$Fe$_{14}$B. 
The results are listed in Table~\ref{table:mag_nd2fe14b}.
From the comparison between Nd$_{2}$Fe$_{14}$B and Nd$_{2}$Fe$_{14}$, we find that the introduction of B decreases both the magnetic moment and magnetization of Nd$_{2}$Fe$_{14}$B.
The magnetovolume effect of B can be evaluated by comparing magnetization of Nd$_{2}$Fe$_{14}$B$_{0}$ with that of Nd$_{2}$Fe$_{14}$. 
The difference between the two comes solely from the structural change. 
The magnetic moment of Nd$_{2}$Fe$_{14}$B$_{0}$ is enhanced by 1\% compared to that of Nd$_{2}$Fe$_{14}$. 
When it is converted to magnetization, however, 
the magnetization is reduced by 0.017 T due to the volume expansion.
By comparing Nd$_{2}$Fe$_{14}$B and Nd$_{2}$Fe$_{14}$B$_{0}$, 
we can see that the chemical effect of B is negative on both the magnetic moment and magnetization. 
The magnitude is larger in the chemical effect than in the magnetovolume effect.
This leads to a decrease of the magnetic moment caused by B. 
This result gives us a different insight from that of the current belief that B enhances the magnetic moment and magnetization of Nd$_{2}$Fe$_{14}$B. 

The local magnetic moments of Fe at each site in Nd$_{2}$Fe$_{14}$B and Nd$_{2}$Fe$_{14}$B$_{0}$ are shown in Fig.\ref{fig:localmoment_nd2fe14b}. 
By following the concept of cobaltization, one can expect the decrease of the Fe magnetic moment at the neighbors of B.
The local magnetic moment at the Fe sites near the pseudo Co, in turn, can be expected to be enhanced due to the presence of pseudo Co.
The Fe atoms at the 16k$_{1}$ and 4e sites are the first and second neighbors of B, and the distances between these Fe atoms and B are about 2.1 \AA. 
Therefore, these Fe atoms become pseudo Co.
In fact, the decrease of the Fe local moments is actually seen at these two sites in Fig.\ref{fig:localmoment_nd2fe14b}. 
All the other Fe sites are near the pseudo Co, thus, we can see the small enhancement of the magnetic moment at these sites.
The reduction at the pseudo Co sites are larger than the increase at the neighbors of the pseudo Co sites in magnitude.
To sum up, B reduces the magnetic moment and magnetization of Nd$_{2}$Fe$_{14}$B.

Next, we examine how the magnetocrystalline anisotropy of Nd$_{2}$Fe$_{14}$B is influenced by the added B. 
To do this, we analyze the crystal-field parameter $A_{2}^{0} \langle r^{2} \rangle$ of Nd. 
The magnetocrystalline anisotropy energy $E_{\textrm {MAE}}$ can be expressed as, 
\begin{equation}
  E_{\text{MAE}} = K_{1} \sin^{2}\theta + K_{2} \sin^{4}\theta + \cdots \;.
\end{equation}
In crystal-field theory, $K_1$ is expressed as, 
\begin{equation}
  K_{1} = -3 J \left(J-\dfrac{1}{2}\right)\Theta_{2}^{0} A_{2}^{0} \langle r^{2} \rangle\;,
\end{equation}
where $A_{2}^{0} \langle r^{2} \rangle$ is a crystal-field parameter. 
The crystal-field parameter of Nd can be calculated from the effective potential obtained through first principles calculations.
Although the magnetocrystalline anisotropy energy includes higher-order contributions,
the magnetocrystalline anisotropy of Nd at high temperatures can be discussed by the lowest order contribution~\cite{Buschow,Sasaki}.
The higher-order crystal-field parameters are discussed in Appendix~\ref{app:crystalfieldparameters}. 
There are two different Wyckoff positions for Nd, labeled as Nd(4f) and Nd(4g).
The calculated $A_{2}^{0} \langle r^{2} \rangle$ are 220 K at Nd(4f) and 330 K at Nd(4g) for Nd$_{2}$Fe$_{14}$B. 
Those for Nd$_{2}$Fe$_{14}$ are 206 K at Nd(4f) and 434 K at Nd(4g).
This result indicates that Nd$_{2}$Fe$_{14}$ already shows uniaxial anisotropy, 
and the presence of B has a minor impact on the magnetocrystalline anisotropy.

From the above results, the advantage of adding B cannot be seen, so far, 
since its addition does not seem to play an important role in terms of enhancing the magnetization and magnetocrystalline anisotropy of Nd$_{2}$Fe$_{14}$B. 
What is the benefit of adding B in Nd$_{2}$Fe$_{14}$B? 
A probable role is to stabilize the structure of Nd$_{2}$Fe$_{14}$B.
In order to confirm that, we examine the stability of Nd$_{2}$Fe$_{14}$B. 
We choose Nd$_{2}$Fe$_{17}$B as a reference system from the historical reason that 
Sagawa intended to insert B to Nd$_{2}$Fe$_{17}$, as explained in Sec.~\ref{introduction}. 
We define the formation energy $\Delta E$ as follows; 
{\setlength\arraycolsep{2pt}
  \begin{eqnarray}
    \Delta E &\equiv& \left(E[\text{Nd}_{2}\text{Fe}_{14}\text{B}] + 3 \mu[\text{Fe}]\right)
    - E[\text{Nd}_{2}\text{Fe}_{17}\text{B}]
    \label{eq:formationenergy_1}
    \\
    &=& \left\{\left(E[\text{Nd}_{2}\text{Fe}_{14}\text{B}] + 3 \mu[\text{Fe}]\right) 
    - \left(E[\text{Nd}_{2}\text{Fe}_{17}] + \mu[\text{B}]\right)\right\}
    \nonumber
    \\
    &-& \left\{E[\text{Nd}_{2}\text{Fe}_{17}\text{B}] 
    - \left(E[\text{Nd}_{2}\text{Fe}_{17}] + \mu[\text{B}]\right)\right\}.
    \label{eq:formationenergy_2}
  \end{eqnarray}
}
$E$[Nd$_{2}$Fe$_{14}$B], $E$[Nd$_{2}$Fe$_{17}$B], $E$[Nd$_{2}$Fe$_{17}$] denote the total energy of Nd$_{2}$Fe$_{14}$B, Nd$_{2}$Fe$_{17}$B and Nd$_{2}$Fe$_{17}$ per formula unit, respectively, 
while $\mu$[Fe] and $\mu$[B] represent chemical potentials corresponding to the total energy of $\alpha$-Fe and $\alpha$-B per atom, respectively. 
Equation~\eqref{eq:formationenergy_2} shows the difference between the formation energies of Nd$_{2}$Fe$_{14}$B and the system in which one B atom is added in Nd$_{2}$Fe$_{17}$ per formula unit. 
The resultant formation energy is $\Delta E = -1.208$ eV/f.u.
From this negative formation energy, we conclude that B in Nd$_{2}$Fe$_{14}$B stabilizes the structure of Nd$_{2}$Fe$_{14}$B. 

\subsection{Comparison with Nd$_{2}$Fe$_{14}X$ ($X$ = C, N, O, F)}
\begin{figure}[t]
\includegraphics[width=8.5cm]{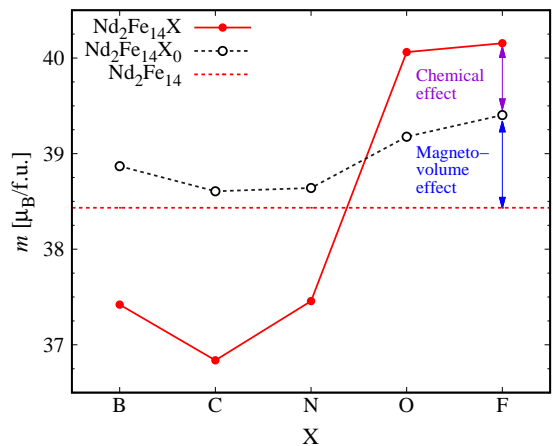}
\caption{(Color online) 
The $X$ dependence of the magnetic moment $m$ [$\mu_{\text{B}}$].
The closed red circles are data for Nd$_{2}$Fe$_{14}X$ and the open black circles are data for Nd$_{2}$Fe$_{14}X_{0}$.
The broken red line illustrates the magnetic moment of Nd$_{2}$Fe$_{14}$.}
\label{fig:magneticmoment_bcnof}
\end{figure}

In this section, we systematically compare the magnetization, magnetic moments, magnetocrystalline anisotropy and stability of Nd$_{2}$Fe$_{14}X$ with the results of Nd$_{2}$Fe$_{14}$B when $X$ is C, N, O and F. 
Figure~\ref{fig:magneticmoment_bcnof} illustrates the magnetic moment of Nd$_{2}$Fe$_{14}X$.
By comparing the broken red line of Nd$_{2}$Fe$_{14}$ 
with the black open circles of Nd$_{2}$Fe$_{14}X{_0}$, we can see that the magnetovolume effect works positively to increase the magnetic moment in all the cases studied here. 
This tendency is proportional to the volume expansion caused by the $X$ elements (see Table~\ref{table:latticeconstant_bcnof} in Appendix~\ref{app:latticeconstants}). 
Similarly, we can evaluate the chemical effects of the $X$ elements on the magnetic moment 
by comparing the black circles with the red circles of Nd$_{2}$Fe$_{14}X$, because the structures of Nd$_{2}$Fe$_{14}X_{0}$ are fixed to those of Nd$_{2}$Fe$_{14}X$.
The difference between these two systems is entirely the absence/presence of the $X$ elements.
We can see that the chemical effect strongly depends on $X$. 
When $X$ = B, C and N, the chemical effect is negative, while it is positive when $X$ = O and F.
Similar trends for the magnetovolume and chemical effects have been reported before in NdFe$_{11}$Ti$X$~\cite{Harashima2} 
(also in a simpler system Fe$_{4}X$~\cite{Akai2}). 
A difference between Nd$_{2}$Fe$_{14}X$ and NdFe$_{11}$Ti$X$ is seen in the total magnetic moment for $X$ = N.
The total magnetic moment is suppressed in the former, whereas it is enhanced in the latter; namely 
NdFe$_{11}$TiN has a larger magnetic moment than NdFe$_{11}$Ti. 

\begin{figure}[bh]
\includegraphics[width=8.5cm]{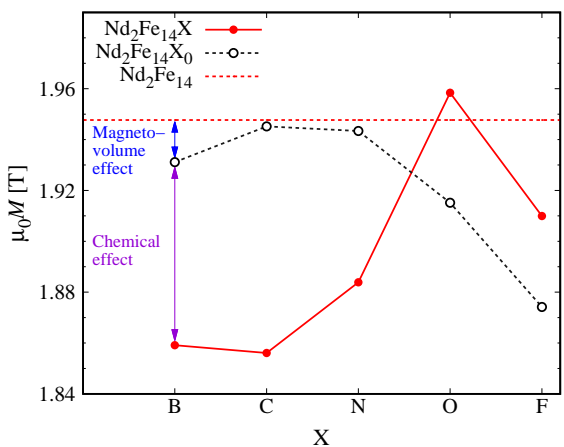}
\caption{(Color online) 
The $X$ dependence of magnetization $\mu_{0}M$ [T].
The closed red circles and the open black circles represent the magnetization of Nd$_{2}$Fe$_{14}X$ and Nd$_{2}$Fe$_{14}X_{0}$, respectively. 
Lines are a visual guide. 
For reference, the broken red line is the magnetization of Nd$_{2}$Fe$_{14}$.}
\label{fig:magnetization_bcnof}
\end{figure}

The above results are converted to the magnetization in Fig.~\ref{fig:magnetization_bcnof}. 
The magnetization is lower in Nd$_{2}$Fe$_{14}X$ than in Nd$_{2}$Fe$_{14}$, except when $X$ = O.
From the comparison between Nd$_{2}$Fe$_{14}$ and Nd$_{2}$Fe$_{14}X_{0}$, 
we can see that the magnetovolume effect is negative in all cases, 
which is in contrast to the positive magnetovolume effect for NdFe$_{11}$Ti$X$~\cite{Harashima2}. 
Especially, the negative effect is large when $X$ is O or F.
On the other hand, the chemical effect is negative when $X$ is B, C or N, but positive when $X$ is O or F. 
The chemical effect plays an important role in enhancing the magnetization of Nd$_{2}$Fe$_{14}$O. 

\begin{figure}[b]
\includegraphics[width=8.5cm]{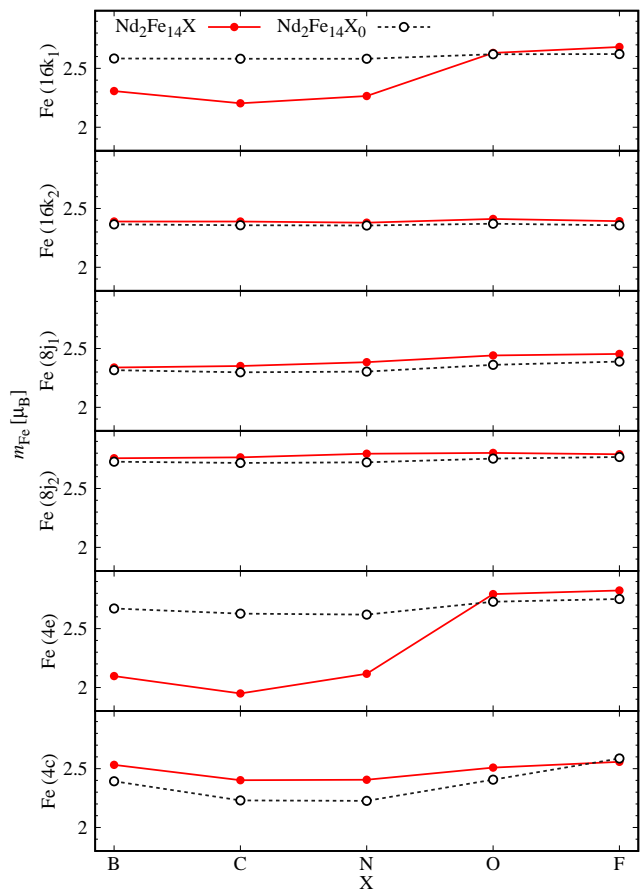}
\caption{
(Color online) The $X$ dependence of the local magnetic moment of Nd$_{2}$Fe$_{14}X$.
The closed red circles are data for Nd$_{2}$Fe$_{14}X$ and the open black circles are data for Nd$_{2}$Fe$_{14}X_{0}$.
The site notation follows Refs.~\onlinecite{Herbst1,Herbst2}.}
\label{fig:localmagneticmoment_bcnof}
\end{figure}
\begin{figure}[t]
\includegraphics[width=8.5cm]{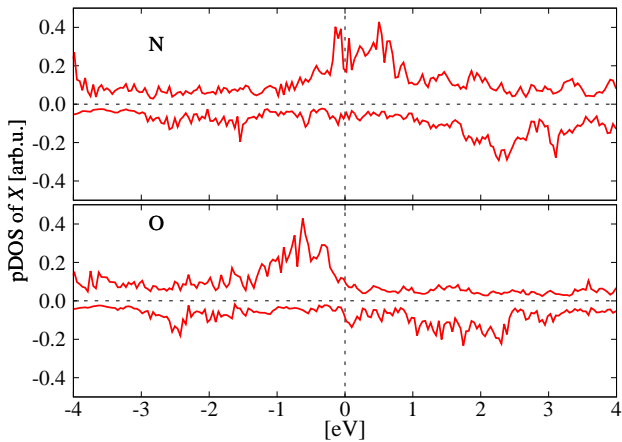}
\caption{(Color online) The local density of states of N in Nd$_{2}$Fe$_{14}$N and O in Nd$_{2}$Fe$_{14}$O.}
\label{fig:dos_x4g_no}
\end{figure}

Figure~\ref{fig:localmagneticmoment_bcnof} illustrates the local magnetic moments at each Fe site in Nd$_{2}$Fe$_{14}X$ and Nd$_{2}$Fe$_{14}X_{0}$. 
We can see the increase of the magnetic moments at 16k$_{1}$ and 4e sites when $X$ changes from N to O.
This enhancement arises from the $p$-$d$ hybridization between O and Fe~\cite{Asano1,Asano2,Harashima2}. 
Figure~\ref{fig:dos_x4g_no} shows the projected density of states on N and O. 
In the case of $X$ = N, there is a peak above the Fermi level in the majority-spin channel.
It comes from the anti-bonding state between N-2$p$ and Fe-3$d$ orbitals. 
The corresponding peak in the minority-spin channel is observed at higher energy (2--3 eV above the Fermi level). 
In the case of $X$ = O, the state is pulled down and is occupied in the majority-spin channel.
This spin-dependent occupancy 
makes a distinction in magnetization between the cases of N and O. 
A similar increase is observed in the local moment of the Fe(8j) sites in NdFe$_{11}$Ti$X$ when $X$ = C and N~\cite{Harashima2}.
The difference can be explained as follows:
In NdFe$_{11}$Ti$X$, the anti-bonding state can be constituted from $\pi$ bonds with surrounding Fe atoms due to the highly symmetric environment at $X$
(even though Ti slightly breaks the symmetry).
In Nd$_{2}$Fe$_{14}X$, one $X$-Fe $\pi$ bond hybridizes with other Fe atoms via $\sigma$ bond.
This $\sigma$ bond is stronger than $\pi$ bond and makes the level of the anti-bonding state higher. 
Consequently, the anti-bonding state is shifted down to the Fermi level and occupied for an element having deeper 2$p$ level, that is, O in Nd$_{2}$Fe$_{14}X$.

\begin{figure}[b]
\includegraphics[width=8.5cm]{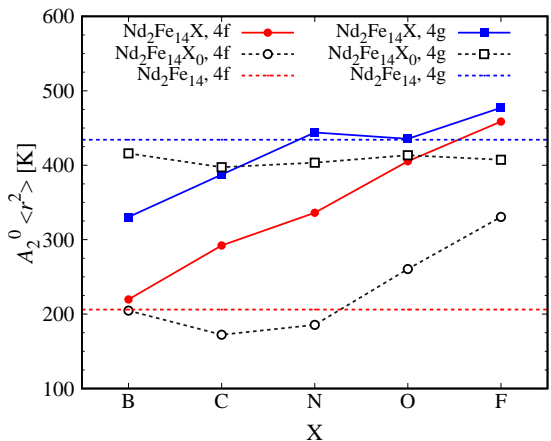}
\caption{\label{fig:A20}
(Color online) $A_{2}^{0} \langle r^{2} \rangle$ of Nd$_{2}$Fe$_{14}X$ and Nd$_{2}$Fe$_{14}X_{0}$.
The closed red circles are data for Nd$_{2}$Fe$_{14}X$ and open black circles for Nd$_{2}$Fe$_{14}X_{0}$. 
The horizontal broken red line is data for Nd$_{2}$Fe$_{14}$.
Lines are a guide to the eye.}
\end{figure}
Figure \ref{fig:A20} illustrates the dependence of the crystal-field parameter $A_{2}^{0} \langle r^{2} \rangle$ of Nd on $X$.
The values of both Nd(4f) and (4g) sites are shown.
They are positive in all cases and $A_{2}^{0} \langle r^{2} \rangle$ increases as the atomic number of $X$ increases (except for Nd(4g) at $X$ = O).
This implies that Nd$_{2}$Fe$_{14}X$ has uniaxial anisotropy at high temperatures and that the anisotropy is enhanced as the atomic number increases.
$A_{2}^{0} \langle r^{2} \rangle$ of Nd(4f) in Nd$_{2}$Fe$_{14}X_{0}$ also increases from $X$ = C to F.
The chemical effect on Nd(4f) is large in magnitude and seems to be similar among $X$ = C, N, O and F.
For Nd(4g) in Nd$_{2}$Fe$_{14}X_{0}$, $A_{2}^{0} \langle r^{2} \rangle$ is almost constant and
$X$ dependence in $A_{2}^{0} \langle r^{2} \rangle$ of Nd(4g) is mainly due to the chemical effect.

\begin{figure}[t]
\includegraphics[width=8.5cm]{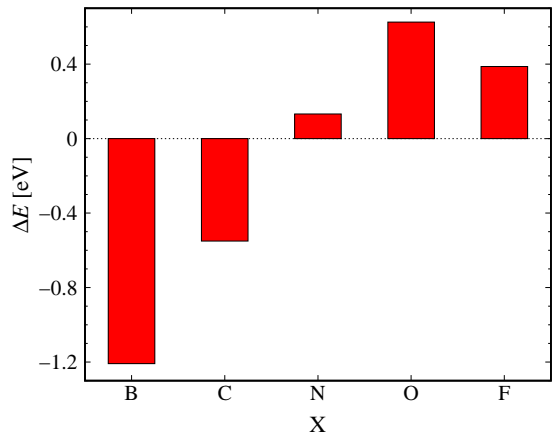}
\caption{(Color online) The formation energy of Nd$_{2}$Fe$_{14}X$ 
defined by Eq.(\ref{eq:formationenergy_1}) with replacement of B with C, N, O, and F.}
\label{fig:formationenergy_bcnof}
\end{figure}

Figure~\ref{fig:formationenergy_bcnof} shows the formation energies of Nd$_{2}$Fe$_{14}X$ as a function of the $X$ elements calculated with Eq.~\eqref{eq:formationenergy_1}, 
but B is replaced with C, N, O and F.
Nd$_{2}$Fe$_{14}X$ is stable when $X$ is B or C, but not when $X$ is N, O or F. 
In other words, Nd$_{2}$Fe$_{17}X$ is more stable when $X$ is N, O or F. 
These results are in good agreement with the fact that Nd$_{2}$Fe$_{14}$B and Nd$_{2}$Fe$_{17}$N$_{x} (0 < x \lesssim 3)$ can be stably synthesized in experiments.
\section{Conclusions}
The roles of added B in Nd$_{2}$Fe$_{14}$B have been carefully investigated by first-principles calculations. 
As reported in previous studies, cobaltized Fe atoms can be seen in Nd$_{2}$Fe$_{14}$B. 
However, magnetization is not enhanced by the added B.
We clarify that both the chemical and magnetovolume effects work negatively on the magnetization of Nd$_{2}$Fe$_{14}$B. 
The magnetocrystalline anisotropy discussed by the crystal-field parameter $A_{2}^{0} \langle r^{2} \rangle$ has a minor effect from the addition of B.
We also systematically examined the effects of the $X$ elements on the magnetism in Nd$_{2}$Fe$_{14}X$ through first-principles calculations. 
The enhancement in the magnetization is observed only for $X$ = O.
The crystal-field parameter $A_{2}^{0} \langle r^{2} \rangle$ has a tendency to increase as the atomic number of $X$ increases. 
The stability of Nd$_{2}$Fe$_{14}X$ was also calculated by comparing the formation energies to that of Nd$_{2}$Fe$_{17}X$. 
The formation energies of Nd$_{2}$Fe$_{14}$B and Nd$_{2}$Fe$_{14}$C are negative relative to Nd$_{2}$Fe$_{17}X$, 
while Nd$_{2}$Fe$_{14}$N, Nd$_{2}$Fe$_{14}$O and Nd$_{2}$Fe$_{14}$F are not energetically stable.
This result is consistent with experimental results. 

\begin{acknowledgments}
We thank Dr. Taro Fukazawa for his fruitful and stimulating discussion.
This work was supported by 
the Elements Strategy Initiative Project under the auspices of MEXT, 
by ``Materials research by Information Integration''
Initiative (MI$^{2}$I) project of the Support Program for Starting Up Innovation Hub from
Japan Science and Technology Agency (JST),
and also by 
MEXT as a social and scientific priority issue 
(Creation of new functional Devices and high-performance Materials to 
Support next-generation Industries; CDMSI) to be tackled by using post-K computer, 
as well as JSPS KAKENHI Grant No.~17K04978.
The calculations were partly carried out by using supercomputers at ISSP, The University of Tokyo, and TSUBAME, Tokyo Institute of Technology, the supercomputer of ACCMS, Kyoto University, and also 
by the K computer, RIKEN (Project No. hp160227, No. hp170100, and No. hp170269).
\end{acknowledgments}

\appendix
\section{Lattice constants}
\label{app:latticeconstants}
The calculated inner coordinates and lattice constants of Nd$_{2}$Fe$_{14}X$ ($X$ = B, C, N, O, F) and Nd$_{2}$Fe$_{14}$ are listed in Table~\ref{table:innercoordinate} and Tables~\ref{table:latticeconstant_bcnof}, respectively.
Volume expansion due to added elements can be seen in all Nd$_{2}$Fe$_{14}X$ systems 
compared to Nd$_{2}$Fe$_{14}$. 
The lattice parameters $c$ become shorter when B is replaced by C and N, and it becomes longer as $X$ = O and F.
This tendency might be related to the $p$-$d$ hybridization between $X$ and the neighboring Fe at the 16k$_{1}$ and 4e sites.
$X$ and the neighboring Fe align along the $c$ axis rather than the $a$ axis and $c$ can reflect the bond length between $X$ and Fe.
From $X$ = B to N, the atomic radius decreases~\cite{Pyykkoe}, thus, the bond length becomes shorter.
\begin{table}[H]
\caption{The inner coordinates for Nd$_{2}$Fe$_{14}X$ and Nd$_{2}$Fe$_{14}$.}
  \vspace{6pt}
  \begin{tabular}{cccccc}
    \hline \hline
    \; compound & \; atom \; & \; site \; & $\qquad x \qquad$ & $\qquad y \qquad$ & $\qquad z \qquad$  \\
    \hline
    \\
                         & Nd & 4f &  0.266 &  0.266 &  0.0  \\
                         &    & 4g &  0.142 & $-$0.142 &  0.0  \\
                         & Fe & 16k$_{1}$ &  0.225 &  0.567 &  0.127  \\
                         &    & 16k$_{2}$ &  0.037 &  0.360 &  0.177  \\
    {Nd$_{2}$Fe$_{14}$B} &    & 8j$_{1}$ &  0.098 &  0.098 &  0.205  \\
                         &    & 8j$_{2}$ &  0.317 &  0.317 &  0.246  \\
                         &    & 4e &  0.5   &  0.5   &  0.114  \\
                         &    & 4c &  0.0   &  0.5   &  0.0  \\
                         & B  & 4g &  0.375 & $-$0.375 &  0.0  \\
    \\
    \hline
    \\
                         & Nd & 4f &  0.257 &  0.257 &  0.0  \\
                         &      & 4g &  0.142 & $-$0.142 &  0.0  \\
                         & Fe & 16k$_{1}$ &  0.225 &  0.566 &  0.120  \\
                         &      & 16k$_{2}$ &  0.036 &  0.359 &  0.177  \\
    {Nd$_{2}$Fe$_{14}$C} &      & 8j$_{1}$ &  0.098 &  0.098 &  0.209  \\
                         &      & 8j$_{2}$ &  0.317 &  0.317 &  0.245  \\
                         &      & 4e &  0.5 &  0.5 &  0.107  \\
                         &      & 4c &  0.0 &  0.5 &  0.0  \\
                         & C  & 4g &  0.374 & $-$0.374 &  0.0  \\
    \\
    \hline
    \\
                         & Nd & 4f &  0.254 &  0.254 &  0.0  \\
                         &      & 4g &  0.142 & $-$0.142 &  0.0  \\
                         & Fe & 16k$_{1}$ &  0.225 &  0.565 &  0.120  \\
                         &      & 16k$_{2}$ &  0.036 &  0.359 &  0.178  \\
    {Nd$_{2}$Fe$_{14}$N} &      & 8j$_{1}$ &  0.098 &  0.098 &  0.209  \\
                         &      & 8j$_{2}$ &  0.317 &  0.317 &  0.244  \\
                         &      & 4e &  0.5 &  0.5 &  0.106  \\
                         &      & 4c &  0.0 &  0.5 &  0.0  \\
                         & N  & 4g &  0.371 & $-$0.371 &  0.0  \\
    \\
    \hline
    \\
                         & Nd & 4f &  0.252 &  0.252 &  0.0  \\
                         &      & 4g &  0.146 & $-$0.146 &  0.0  \\
                         & Fe & 16k$_{1}$ &  0.224 &  0.562 &  0.129  \\
                         &      & 16k$_{2}$ &  0.035 &  0.360 &  0.177  \\
    {Nd$_{2}$Fe$_{14}$O} &      & 8j$_{1}$ &  0.098 &  0.098 &  0.204  \\
                         &      & 8j$_{2}$ &  0.316 &  0.316 &  0.244  \\
                         &      & 4e &  0.5 &  0.5 &  0.117  \\
                         &      & 4c &  0.0 &  0.5 &  0.0  \\
                         & O  & 4g &  0.363 & $-$0.363 &  0.0  \\
    \\
    \hline
    \\
                         & Nd & 4f &  0.258 &  0.258 &  0.0  \\
                         &      & 4g &  0.146 & $-$0.146 &  0.0  \\
                         & Fe & 16k$_{1}$ &  0.224 &  0.563 &  0.141  \\
                         &      & 16k$_{2}$ &  0.036 &  0.362 &  0.174  \\
    {Nd$_{2}$Fe$_{14}$F} &      & 8j$_{1}$ &  0.098 &  0.098 &  0.197  \\
                         &      & 8j$_{2}$ &  0.317 &  0.317 &  0.246  \\
                         &      & 4e &  0.5 &  0.5 &  0.135  \\
                         &      & 4c &  0.0 &  0.5 &  0.0  \\
                         & F  & 4g &  0.369 & $-$0.369 &  0.0  \\
    \\
    \hline
    \\
                         & Nd & 4f &  0.256 &  0.256 &  0.0  \\
                         &      & 4g &  0.139 & $-$0.139 &  0.0  \\
                         & Fe & 16k$_{1}$ &  0.229 &  0.566 &  0.120  \\
    {Nd$_{2}$Fe$_{14}$} &      & 16k$_{2}$ &  0.035 &  0.360 &  0.177  \\
                         &      & 8j$_{1}$ &  0.098 &  0.098 &  0.210  \\
                         &      & 8j$_{2}$ &  0.316 &  0.316 &  0.245  \\
                         &      & 4e &  0.5   &  0.5   &  0.103  \\
                         &      & 4c &  0.0   &  0.5   &  0.0  \\
    \hline \hline
  \end{tabular}
  \label{table:innercoordinate}
\end{table}
\begin{table}[t]
\caption{
The calculated lattice constants and volumes of Nd$_{2}$Fe$_{14}X$ and Nd$_{2}$Fe$_{14}$ (denoted as E). 
The lattice constants $a$ and $c$ are for a conventional unit cell described in \AA.
The volumes $V$ are per formula unit described in \AA$^{3}$.}
  \vspace{5pt}
  \begin{tabular}{lcccccc}
    \hline \hline
    $X$ = & B & C & N & O & F & E
    \\
    \hline
    $a$ [\AA] & 8.791 & 8.805 & 8.816 & 8.858 & 8.833 & 8.760
    \\
    $c$ [\AA] & 12.141 & 11.934 & 11.925 & 12.154 & 12.561 & 11.989
    \\
    $V$ [\AA$^{3}$/f.u.] & 234.6 & 231.3 & 231.7 & 238.4 & 245.0 & 230.0
    \\
    \hline \hline
  \end{tabular}
  \label{table:latticeconstant_bcnof}
\end{table}
At $X$ = O, the anti-bonding state starts to be occupied and the occupation weakens the covalent bond between $X$ and Fe.
As a result, $c$ in Nd$_{2}$Fe$_{14}$O and Nd$_{2}$Fe$_{14}$F become longer. 
Analogous trend is also observed in NdFe$_{11}$Ti$X$~\cite{Harashima2}.

\section{Crystal-field parameters}
\label{app:crystalfieldparameters}
In this appendix, we discuss the crystal-field parameters $A_{l}^{m} \langle r^{l} \rangle$ of Nd.
The large magnetocrystalline anisotropy of Nd$_{2}$Fe$_{14}$B is attributed to well-localized Nd-4$f$ electrons with large orbital moments.
In crystal-field theory, the magnetocrystalline anisotropy energy $E_{\mathrm{MAE}}(\theta,\varphi)$ of Nd in Nd$_{2}$Fe$_{14}X$ reads 
from the crystal-field hamiltonian $\mathcal{H}_{\text{CEF}}$;
\begin{eqnarray}
  E_{\mathrm{MAE}} (\theta,\varphi) &=& \langle J, M=-J \left| \mathcal{H}_{\text{CEF}} \right| J, M=-J \rangle
  \label{eq:mae_cef_1}
  \\
  &=& \sum_{lm} \Theta_{l} A_{l}^{m} \langle r^{l} \rangle f_{l}(J) \dfrac{Z_{l}^{m}(\theta,\varphi)}{a_{lm}},
  \label{eq:mae_cef_2}
\end{eqnarray}
where 
\begin{eqnarray}
f_{2} &=& J\left(J-\dfrac{1}{2}\right) \nonumber,
  \\
f_{4} &=& J\left(J-\dfrac{1}{2}\right)\left(J-1\right)\left(J-\dfrac{3}{2}\right),
  \\
f_{6} &=& J\left(J-\dfrac{1}{2}\right)(J-1)\left(J-\dfrac{3}{2}\right)(J-2)\left(J-\dfrac{5}{2}\right) \nonumber, 
  \\
  \nonumber
\end{eqnarray}
\begin{eqnarray}
\Theta_{2} &=& - \dfrac{7}{3^{2} \cdot 11^{2}} \nonumber,
  \\
\Theta_{4} &=& - \dfrac{2^{3} \cdot 17}{3^{3} \cdot 11^{3} \cdot 13},
  \\
\Theta_{6} &=& - \dfrac{5 \cdot 17 \cdot 19}{3^{3} \cdot 7 \cdot 11^{3} \cdot 13^{2}} \nonumber,
  \\
  \nonumber
\end{eqnarray}
\begin{eqnarray}
a_{2 -2} &=& a_{2 2} = \dfrac{1}{4}\sqrt{\dfrac{15}{\pi}}  \nonumber,
  \\
a_{2 0} &=& \dfrac{1}{4}\sqrt{\dfrac{5}{\pi}}  \nonumber,
  \\
a_{4 -4} &=& a_{4 4} = \dfrac{3}{16}\sqrt{\dfrac{35}{\pi}} \nonumber,
  \\
a_{4 -2} &=& a_{4 2} = \dfrac{3}{8}\sqrt{\dfrac{5}{\pi}} \nonumber,
  \\
a_{4 0} &=& \dfrac{3}{16}\sqrt{\dfrac{1}{\pi}},
  \\
a_{6 -6} &=& a_{6 6} = \dfrac{231}{64}\sqrt{\dfrac{26}{231\pi}} \nonumber,
  \\
a_{6 -4} &=& a_{6 4} = \dfrac{21}{32}\sqrt{\dfrac{13}{7\pi}} \nonumber,
  \\
a_{6 -2} &=& a_{6 2} = \dfrac{1}{64}\sqrt{\dfrac{2730}{\pi}} \nonumber,
  \\
a_{6 0} &=& \dfrac{1}{32}\sqrt{\dfrac{13}{\pi}} \nonumber.
\end{eqnarray}
This formulation is referred to Ref.~\onlinecite{Yamada}.
Constant parameters are the Stevens factor $\Theta_{l}$ and $f_{l}$ in which the total angular momentum $J$ $(=L + S)$ is 9/2 $(L = 6, S = 3/2)$ for Nd. 
$Z_{l}^{m}$ is a real spherical harmonics.
$A_{l}^{m} \langle r^{l} \rangle$ are crystal-field parameters at the Nd site. 
Here $l$ takes 2, 4, or 6, and $m$ is even number from $-l$ to $l$.
\begin{equation} \label{eq:Alm}
  A_{l}^{m} \langle r^{l} \rangle = a_{lm}\int_{0}^{r_{c}} dr \; r^{2} |R_{4f}(r)|^{2} V_{lm}(r).
\end{equation}
We use the effective potential $V_{lm}(r)$ of Nd atoms obtained by solving the Kohn-Sham equation self-consistently. 
$r_{c}$ is a cutoff radius which is determined by Bader analysis as in Ref.~\onlinecite{Harashima2}, 
and $R_{4f}(r)$ is an atomic radial function of the Nd-4$f$ electrons.
The results for $A_{l}^{m} \langle r^{l} \rangle$ of Nd at the 4f and 4g sites in Nd$_{2}$Fe$_{14}X$ and Nd$_{2}$Fe$_{14}$ are summarized in Table~\ref{table:alm_bcnof}.
The comparison with the previous calculation~\cite{Hummler} and experiment~\cite{Yamada} is also shown for Nd$_{2}$Fe$_{14}$B.
The order of magnitude for $l$ = 2 and 4 agrees with the previous calculated values, 
and the sign for them also agrees, except for $(l,m)=(4,-2)$ of Nd(4g).

\begin{figure}[H]
\includegraphics[width=8.5cm]{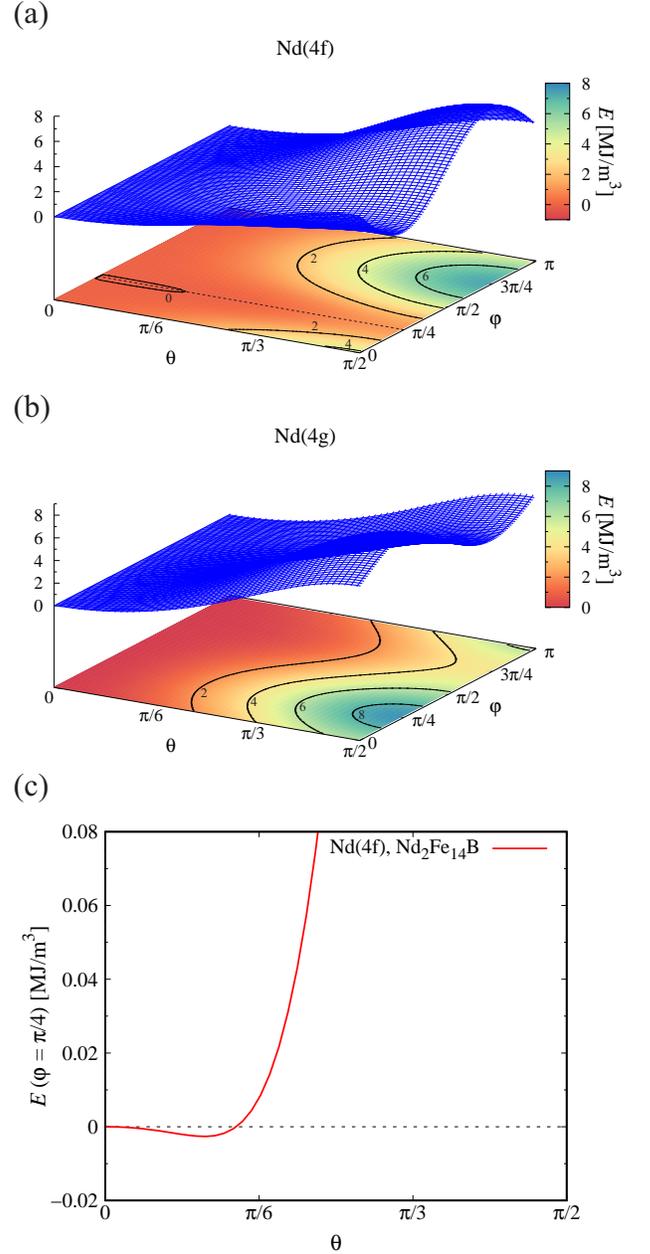}
\caption{(Color online) The spin angle dependence of MAE for (a) Nd(4f) and (b) Nd(4g) in Nd$_{2}$Fe$_{14}$B. 
The solid lines with numbers in the color map represent the contour lines of MAE. 
The dashed line is a guide to the eye for the data in (c).
(c) The angle dependence of MAE for Nd(4f) in Nd$_{2}$Fe$_{14}$B at $\varphi$ = $\pi/4$.
}
\label{fig:mae-angle_4f_4g}
\end{figure}
Figures~\ref{fig:mae-angle_4f_4g} (a) and (b) show the angle dependence of Eq.~\eqref{eq:mae_cef_2} for Nd(4f) and Nd(4g) in Nd$_{2}$Fe$_{14}$B.
Strong dependence on the azimuthal angle $\varphi$ can be seen in the MAE for both Nd(4f) and Nd(4g). 
Figure~\ref{fig:mae-angle_4f_4g} (c) shows the MAE at $\varphi$ = $\pi/4$ for Nd(4f).
The negative region appears around $\theta$ = 0.
We here note that the exchange coupling with Fe is not taken into account 
and that such small contribution of MAE may not survive in the presence of the strong exchange coupling.
Therefore, we cannot conclude the existence of the spin canting at low temperatures 
from the present result~\cite{HiMaYaFuSaYa1986}. 
However, it indicates the importance of the azimuthal angle dependence of the MAE to describe the spin configuration at low temperature. 

\begin{table*}
\caption{
The calculated $A_{l}^{m} \langle r^{l} \rangle$ of Nd at the 4f and 4g sites 
in Nd$_{2}$Fe$_{14}X$ (= B, C, N, O, F) and Nd$_{2}$Fe$_{14}$.
They are described in Kelvin.}
  \vspace{5pt}
  \begin{tabular}{lccccccccc}
    \hline \hline
    & \; $A_{2}^{0} \langle r^{2} \rangle$ \; 
    & \; $A_{2}^{-2} \langle r^{2} \rangle$ \; 
    & \; $A_{4}^{0} \langle r^{4} \rangle$ \; 
    & \; $A_{4}^{-2} \langle r^{4} \rangle$ \; 
    & \; $A_{4}^{4} \langle r^{4} \rangle$ \; 
    & \; $A_{6}^{0} \langle r^{6} \rangle$ \; 
    & \; $A_{6}^{-2} \langle r^{6} \rangle$ \; 
    & \; $A_{6}^{4} \langle r^{6} \rangle$ \; 
    & \; $A_{6}^{-6} \langle r^{6} \rangle$ \; 
    \\
    \hline
    Nd(4f) &&&&&&&&&
    \\
    Nd$_{2}$Fe$_{14}$B & 220 &  442 & -21 & -17 & -18 & 2.6 &  1.5 & -3.0 &  0.0
    \\
    Nd$_{2}$Fe$_{14}$C & 292 &  579 & -23 & -54 & -37 & 4.1 &  7.9 &  2.9 &  4.7
    \\
    Nd$_{2}$Fe$_{14}$N & 336 &  627 & -26 & -73 & -50 & 4.2 & 10.0 &  1.9 & 10.1
    \\
    Nd$_{2}$Fe$_{14}$O & 405 &  652 & -28 & -46 & -59 & 4.5 &  6.5 &  7.4 & 15.3
    \\
    Nd$_{2}$Fe$_{14}$F & 459 &  691 & -23 & -12 & -45 & 4.0 & -1.1 & 10.6 & 15.9
    \\
    Nd$_{2}$Fe$_{14}$  & 206 &  372 & -19 & -71 & -34 & 4.3 &  7.9 & -1.0 & 11.4
    \\
    \\
    Nd(4g) &&&&&&&&&
    \\
    Nd$_{2}$Fe$_{14}$B & 331 & -327 & -30 & -28 &  34 & 3.4 & -3.2 &  3.0 & 11.9
    \\
    Nd$_{2}$Fe$_{14}$C & 387 & -431 & -33 & -34 &  47 & 3.5 &  1.5 & 10.2 &  0.8
    \\
    Nd$_{2}$Fe$_{14}$N & 444 & -532 & -36 & -36 &  58 & 2.9 &  4.2 & 14.1 & -4.0
    \\
    Nd$_{2}$Fe$_{14}$O & 435 & -562 & -32 & -70 & -13 & 1.9 &  5.1 &  6.4 &  0.7
    \\
    Nd$_{2}$Fe$_{14}$F & 477 & -658 & -28 & -68 &  33 & 2.0 &  6.0 &  7.7 &  6.8
    \\
    Nd$_{2}$Fe$_{14}$  & 434 & -308 & -34 &  23 &  16 & 3.2 &  2.6 &  8.1 & -0.3
    \\
    \hline \hline
  \end{tabular}
  \label{table:alm_bcnof}
\end{table*}

\begin{table*}
\caption{
  The comparison of the calculated $A_{l}^{m} \langle r^{l} \rangle$ with previous theoretical~\cite{Hummler,Yoshioka} and experimental~\cite{Yamada} studies for Nd$_{2}$Fe$_{14}$B. 
  They are described in Kelvin.
  In Ref.~\onlinecite{Hummler}, the notation of the Nd sites, 4f and 4g, are opposite from the present notation, 
  and here it is alternated to compare with our data.
  The data for Nd1 and Nd2 on the middle layer described in Ref.~\onlinecite{Yoshioka} correspond to Nd(4f) and Nd(4g), respectively.
  The data in Ref.~\onlinecite{Yamada} is shown as $A_{l}^{m}$ [K$a_{\text{B}}^{-l}$] itself, and 
  we multiply $\langle r^{2} \rangle$ = 1.001 $a_{\text{B}}^{2}$, $\langle r^{4} \rangle$ = 2.401 $a_{\text{B}}^{4}$, and $\langle r^{6} \rangle$ = 12.396 $a_{\text{B}}^{6}$,
  to the values of $A_{l}^{m}$, respectively~\cite{Freeman}.
  Nd(4f) and Nd(4g) sites are not distinguished in Ref.~\onlinecite{Yamada}.
  For $A_{2}^{-2} \langle r^{2} \rangle$ and $A_{6}^{-6} \langle r^{6} \rangle$, 
  the values at Nd(4f) and Nd(4g) on one plane is the same magnitude of each Nd on the plane shifted by 0.5$c$ with opposite signs.
}
  \vspace{5pt}
  \begin{tabular}{llccccccccc}
    \hline \hline
    &
    & \; $A_{2}^{0} \langle r^{2} \rangle$ \; 
    & \; $A_{2}^{-2} \langle r^{2} \rangle$ \; 
    & \; $A_{4}^{0} \langle r^{4} \rangle$ \; 
    & \; $A_{4}^{-2} \langle r^{4} \rangle$ \; 
    & \; $A_{4}^{4} \langle r^{4} \rangle$ \; 
    & \; $A_{6}^{0} \langle r^{6} \rangle$ \; 
    & \; $A_{6}^{-2} \langle r^{6} \rangle$ \; 
    & \; $A_{6}^{4} \langle r^{6} \rangle$ \; 
    & \; $A_{6}^{-6} \langle r^{6} \rangle$ \; 
    \\
    \hline
    present              & Nd(4f) & 220 &  442 & -21   & -17 &  -18 &   2.6 &   1.5 &  -3.0 &  0.0
    \\
                         & Nd(4g) & 331 & -327 & -30   & -28 &   34 &   3.4 &  -3.2 &   3.0 & 11.9
    \\
    \\
    calc.~\cite{Hummler}  & Nd(4f) & 323 &  807 & -50   & -85 & -117 &  -2.7 &   3.9 &  -24  &  0.1
    \\
                         & Nd(4g) & 540 & -727 & -54   &  82 &  125 &  -2.2 &   0.3 &  -15  &  5
    \\
    \\
    calc.~\cite{Yoshioka} & Nd(4f) & 442 &  777 & -28   & -69 & -106 &  17   &  -2   &  156  & 57
    \\
                         & Nd(4g) & 605 & -286 & -35   &  84 &   96 &  13   &   2   &   79  & 11
    \\
    \\
    exp.~\cite{Yamada}    &        & 295 & -454 & -29.5 &     &      & -22.8 & 121   & -197  & 
    \\
    \hline \hline
  \end{tabular}
  \label{table:alm_comparison}
\end{table*}

\bibliography{NdFeX}

\end{document}